\documentclass[iop,apj,twocolappendix]{emulateapj}
\usepackage{amsmath,amssymb,amstext}

\usepackage[breaklinks,colorlinks,citecolor=blue,linkcolor=magenta]{hyperref} 

\usepackage[all]{hypcap} 
\usepackage{multirow}
\usepackage{booktabs}
\usepackage{enumitem}
\usepackage{rotating}

\usepackage{aas_macros}
\usepackage{natbib}
\usepackage[percent]{overpic}
\usepackage{xcolor}

\usepackage{lineno}
\shorttitle{Star Formation Histories of Ultra-Faint Dwarf Galaxies}

\shortauthors{Sacchi et al.}

\begin{document}

\title{Star Formation Histories of Ultra-Faint Dwarf Galaxies: environmental differences between Magellanic and non-Magellanic satellites? \footnotemark[$\star$]} \footnotetext[$\star$]{Based on observations obtained with the NASA/ESA \textit{Hubble Space Telescope} at the Space Telescope Science Institute, which is operated by the Association of Universities for Research in Astronomy under NASA Contract NAS 5-26555.}
\author{Elena Sacchi$^{1,2,3}$,
Hannah Richstein$^{4}$,
Nitya Kallivayalil$^{4}$,
Roeland van der Marel$^{1,5}$,
Mattia Libralato$^{6}$,
Paul Zivick$^{4}$,
Gurtina Besla$^{7}$,
Thomas M. Brown$^{1}$,
Yumi Choi$^{1}$,
Alis Deason$^{8,9}$,
Tobias Fritz$^{10,11}$,
Marla Geha$^{12}$,
Puragra Guhathakurta$^{13}$,
Myoungwon Jeon$^{14}$,
Evan Kirby$^{15,16}$,
Steven R. Majewski$^{4}$,
Ekta Patel$^{17,18}$,
Joshua D. Simon$^{19}$,
Sangmo Tony Sohn$^{1}$,
Erik Tollerud$^{1}$, and
Andrew Wetzel$^{20}$
}
\affil{
$^{1}$Space Telescope Science Institute, 3700 San Martin Drive, Baltimore, MD 21218, USA\\
$^{2}$ Leibniz-Institut für Astrophysik Potsdam, An der Sternwarte 16, 14482 Potsdam, Germany; esacchi@aip.de\\
$^{3}$INAF--Osservatorio di Astrofisica e Scienza dello Spazio di Bologna, Via Gobetti 93/3, I-40129 Bologna, Italy \\
$^{4}$University of Virginia, Department of Astronomy, 530 McCormick Road, Charlottesville, VA 22904, USA \\
$^{5}$Center for Astrophysical Sciences, Department of Physics \& Astronomy, Johns Hopkins University, Baltimore, MD 21218, USA \\
$^{6}$AURA for the European Space Agency (ESA), ESA Office, Space Telescope Science Institute, 3700 San Martin Drive, Baltimore, MD 21218, USA \\
$^{7}$Steward Observatory, University of Arizona, 933 North Cherry Avenue, Tucson, AZ 85721-0065, USA \\
$^{8}$Institute for Computational Cosmology, Department of Physics, University of Durham, South Road, Durham DH1 3LE, UK \\
$^{9}$Centre for Extragalactic Astronomy, Department of Physics, University of Durham, South Road, Durham DH1 3LE, UK \\
$^{10}$Instituto de Astrofísica de Canarias, Calle Via Láctea s/n, 38206, La Laguna, Tenerife, Spain \\
$^{11}$Universidad de La Laguna (ULL), Departamento de Astrofísica, 30206, La Laguna, Tenerife, Spain \\
$^{12}$Department of Astronomy, Yale University, 52 Hillhouse Ave., New Haven, CT 06520, USA \\
$^{13}$UCO/Lick Observatory, Department of Astronomy \& Astrophysics, University of California Santa Cruz, 1156 High Street, Santa Cruz, CA 95064, USA \\
$^{14}$School of Space Research, Kyung Hee University, 1732 Deogyeong-daero, Yongin-si, Gyeonggi-do 17104, Korea \\
$^{15}$Department of Astronomy, California Institute of Technology, 1200 E California Boulevard, Pasadena, CA 91125, USA \\
$^{16}$Department of Physics, University of Notre Dame, Notre Dame, IN 46556, USA \\
$^{17}$Department of Astronomy, University of California, Berkeley, 501 Campbell Hall, Berkeley, CA, 94720, USA \\
$^{18}$Miller Institute for Basic Research in Science, 468 Donner Lab, Berkeley, CA 94720, USA \\
$^{19}$Observatories of the Carnegie Institution for Science, 813 Santa Barbara Street, Pasadena, CA 91101, USA \\
$^{20}$Department of Physics \& Astronomy, University of California, Davis, CA 95616, USA
}

%TC:ignore
\begin{abstract}
We present the color-magnitude diagrams and star formation histories (SFHs) of seven ultra-faint dwarf galaxies: Horologium~1, Hydra~2, Phoenix~2, Reticulum~2, Sagittarius~2, Triangulum~2, and Tucana~2, derived from high-precision \textit{Hubble Space Telescope} photometry. We find that the SFH of each galaxy is consistent with them having created at least 80\% of the stellar mass by $z\sim6$. For all galaxies, we find quenching times older than 11.5 Gyr ago, compatible with the scenario in which reionization suppresses the star formation of small dark matter halos. However, our analysis also reveals some differences in the SFHs of candidate Magellanic Cloud satellites, i.e., galaxies that are likely satellites of the Large Magellanic Cloud and that entered the Milky Way potential only recently. Indeed, Magellanic satellites show quenching times about 600 Myr more recent with respect to those of other Milky Way satellites, on average, even though the respective timings are still compatible within the errors. This finding is consistent with theoretical models that suggest that satellites' SFHs may depend on their host environment at early times, although we caution that within the error bars all galaxies in our sample are consistent with being quenched at a single epoch.
\end{abstract}

\keywords{galaxies: ultra faint dwarf -- galaxies: evolution -- galaxies: star formation -- galaxies: stellar content -- galaxies: kinematics and dynamics -- Local Group -- Magellanic Clouds}
%TC:endignore

\maketitle

\section{Introduction}
Ultra-faint dwarf (UFD) galaxies are interesting and peculiar objects, representing many extremes in terms of galaxy properties. Their population includes the least luminous, least chemically-enriched, most dark matter (DM) dominated, and oldest satellite galaxies of the Milky Way (MW; \citealt{Simon2019}). UFDs could be the relics of the first galaxies believed to form, and therefore provide us with a fossil record of the conditions for star formation in the era of reionization (\citealt{Bovill2009} and references therein). According to the prevailing Lambda Cold Dark Matter ($\Lambda$CDM) cosmological scenario, an important channel for mass growth of DM halos is hierarchical accretion. Indeed, simulations show that even low-mass host halos have substructures down to their resolution limit \citep{Wetzel2016,Dooley2017,Besla2018,Jahn2019,Wang2020}. UFDs would also undergo hierarchical growth, and are thus precious tools to study the physics of galaxy assembly in the early Universe.

Recent simulations have illustrated in detail that star formation (SF) in UFDs is impacted by the local background UV ionizing field and stellar feedback, with both acting as effective quenching mechanisms at such low-mass scales \cite[e.g.,][]{Jeon2017,Jeon2021,Wheeler2019,Applebaum2021}. The timescales in these simulations are such that shortly after the epoch of reionization ends ($z\sim6$), UFDs are effectively quenched, although the exact duration of SF can vary depending on the halo mass, even for a fixed ionization background \citep{Jeon2017}, and often some residual interstellar medium remains in the galaxies and can fuel SF for another 1 - 2 Gyr \citep{Wheeler2019}. Analyses of zoom-in simulations of UFDs in MW-like environments have found that traditional environmental effects (e.g., tidal field, ram pressure) are not primary factors in the quenching timescale \citep{Applebaum2021}.

Observational data support a ubiquitous quenching timescale for UFDs around the time of reionization \citep{Brown2014,Weisz2014}. Given that the MW accreted these systems at different times \citep{Fritz2018}, this indicates a global rather than a local physical explanation for the common quenching timescale. However, other studies \citep[e.g.,][]{Joshi2021} have shown a dependence of the quenching timescale on host mass, presumably because the strength of the background and local UV field changes depending on the environment (more massive hosts would have more SF and thus produce more ionizing radiation). However, this study explores the mass range $10^{7-10}$~M$_{\odot}$, thus their results might not hold for the lower-mass galaxies we are interested in.

UFDs are hard to identify due to their low luminosities $M_V > -8$, implying stellar masses $M_{\star} \lesssim 10^4$~M$_{\odot}$, and generally old ($>10$~Gyr) stellar populations \citep{Simon2019}, which lack bright young stars that would ease their discovery and identification. However, a great effort has been made in the past few years to increase the statistics of satellites around the MW, and many new UFDs were discovered thanks to wide-field surveys such as PAN-STARRS \citep{Laevens2015}, DES \citep{Bechtol2015,Drlica-Wagner2015,Koposov2015}, and ATLAS \citep{Torrealba2016}.

Many of these new satellites were found in the proximity of the Magellanic Clouds (MC), a region targeted by several deep imaging surveys (e.g., \citealt{Drlica-Wagner2015,Martin2015,Nidever2017,Koposov2018,Torrealba2018}). These surveys provide a great opportunity to test the self-similarity of $\Lambda$CDM, which predicts that MW satellites, such as the Large Magellanic Cloud (LMC), should also have their own satellites, which fell into the MW potential with the whole Magellanic system \cite[e.g.,][]{Donghia2008,Sales2011,Deason2015,Jahn2019}. One way to test association is to reconstruct their 3D kinematics and orbital history, which is now possible thanks to detailed proper motion (PM) measurements enabled by the Gaia mission. 

\begin{table*}[ht]
\caption{Properties of the seven UFD galaxies analyzed here. Columns 1 and 2 list the galaxy names, columns 3 and 4 the galactocentric coordinates, columns 5 and 6 the distance from the Sun (in kpc) and distance modulus (\citealt{Fritz2018,Fritz2019}, and references therein), column 7 the number of stars used in this study, column 8 the estimated stellar mass (\citealt{Sales2017} and references therein; for cases without estimates of M$_{\star}$, it was derived from the listed $V$-band magnitude assuming a mass-to-light ratio $\Upsilon=2$ in solar units), column 9 the $V$-band magnitude, column 10 the $V$-band extinction, and column 11 the average metallicity (\citealt{Fritz2018,Fritz2019}, and references therein). The last column indicates whether a galaxy is a potential Magellanic satellite, according to the analysis in \citet{Erkal2019} and \citet{Patel2020}.}
\begin{center}
%\resizebox{\columnwidth}{!}{
\begin{tabular}{lcrrrcrccccc}
    \toprule
    \midrule
    \addlinespace[0.3em]
    Galaxy & Abbreviation & $l$ [deg] & $b$ [deg] & d [kpc] & DM & N$_{\star}$ & M$_{\star}$ [M$_{\odot}$] & $M_V$ & $A_V$ & $\langle$[Fe/H]$\rangle$ &  MC satellite? \\
    \midrule
    Horologium 1 & Hor 1 & 270.9	& $-$54.7	& 83.2	& 19.60 & 483 & $1.96 \times 10^3$ & $-$3.4 & 0.04 & $-$2.76 & yes \\
    Hydra 2 & Hya 2 & 295.6	& 30.5	& 150.7 & 20.89 & 334 & $7.10 \times 10^3$ & $-$4.8 & 0.17 & $-$2.02 & \\
    Phoenix 2 & Phx 2 & 323.3	& $-$60.2 & 80.0	& 19.52	& 284 & $1.13 \times 10^3$ & $-$2.8 & 0.03 & $-$2.51 & yes \\
    Reticulum 2 & Ret 2 & 265.9	& $-$49.6	& 31.6	& 17.50 & 237 & $1.00 \times 10^3$ & $-$2.7 & 0.05 & $-$2.46 & yes \\
    Sagittarius 2 & Sag 2 & 18.9	& $-$22.9	& 66.1	& 19.10 & 2199 & $2.47 \times 10^3$ & $-$5.2 &  0.34 & $-$2.81 & \\
    Triangulum 2 & Tri 2 & 140.9	& $-$23.8	& 28.4	& 17.27 & 237 & $8.97 \times 10^2$ & $-$1.8 & 0.22 & $-$2.38 & \\
    Tucana 2 & Tuc 2 & 327.9	& -52.8	& 57.5	& 18.80 & 158 & $4.90 \times 10^3$ & $-$3.8 & 0.05 & $-$2.23 & \\
    \midrule
    \bottomrule
\end{tabular}
%}
\end{center}
\label{catalog}
\end{table*}

\citet{Kallivayalil2018} analyzed the PMs and radial velocities of 13 UFDs using Gaia data release (DR) 2, compared their kinematics to the tidal debris of a simulated analog of the LMC, and found four UFDs whose kinematics are compatible with the LMC debris, Carina 2, Carina 3, Horologium 1, and Hydrus 1. Using a different technique, \citet{Erkal2019} used Gaia DR2 PMs to rewind the satellite orbits from their present day positions and determine which ones were originally bound to the LMC; of the 25 analyzed UFDs they concluded that six, Carina 2, Carina 3, Horologium 1, Hydrus 1, Reticulum 2, and Phoenix 2, are highly compatible with a Magellanic origin. Recently, \citet{Patel2020} calculated the orbital histories of 13 ultra-faint satellites including the combined potential of the MW, LMC, and Small Magellanic Cloud (SMC) for the first time; in this scenario, they identified Carina 2, Carina 3, Horologium 1, and Hydrus 1 as long-term Magellanic satellites, and Reticulum 2 and Phoenix 2 as recently captured Magellanic satellites.

Although these works provide fundamental contributions to our understanding of the MW satellites' dynamics, we need star formation histories (SFHs) to fully explore their properties and understand the impact of environment and reionization on such low mass galaxies. Within the Local Group, the \textit{Hubble Space Telescope} (\textit{HST}) is able to resolve individual stars in galaxies down to several magnitudes below the oldest main sequence (MS) turnoff, allowing us to measure their ancient SFHs and explore differences in the SF quenching behavior \cite[as in, e.g.,][]{Brown2014}. A detailed analysis and comparison of UFDs residing in different environments at early times is also one way to discover variations in the ionization field over large scales.

Here we present an analysis of the optical color-magnitude diagrams (CMDs) and SFHs of seven UFD galaxies part of the \textit{HST} Treasury Program 14734 (PI: Kallivayalil). They are listed in Table \ref{catalog}, together with their distances, in the range $\sim 30 - 150$~kpc, $V$-band magnitudes, between $-1.8$ and $-5.2$, and possible association with the Magellanic Clouds. Our analysis is based on high-precision photometry from the Advanced Camera for Surveys (ACS) and literature spectroscopic measurements, and employs the synthetic CMD method to derive the SFH of each galaxy. \\

\begin{figure*}
\centering
\includegraphics[width=\linewidth]{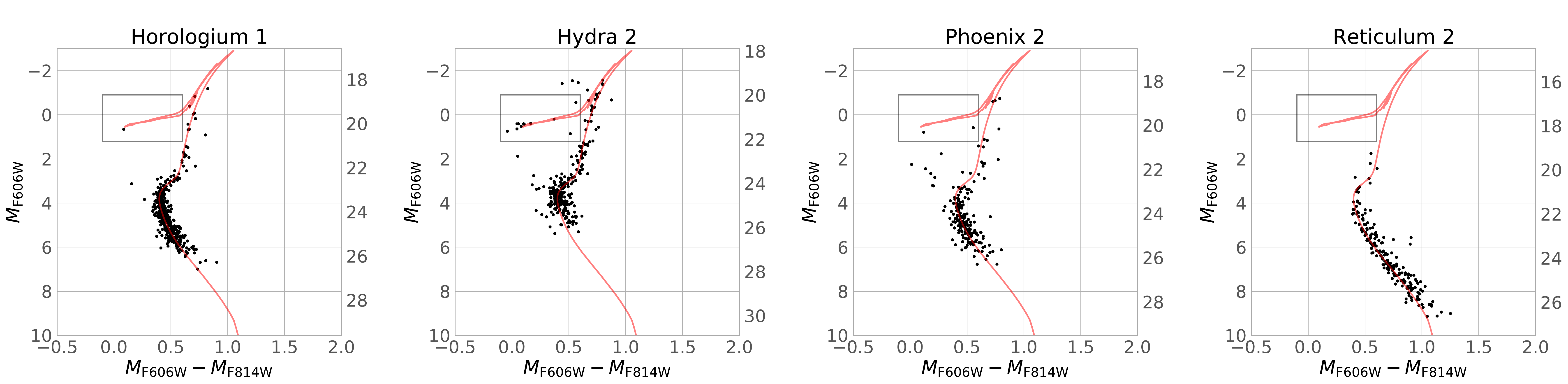}
\includegraphics[width=0.75\linewidth]{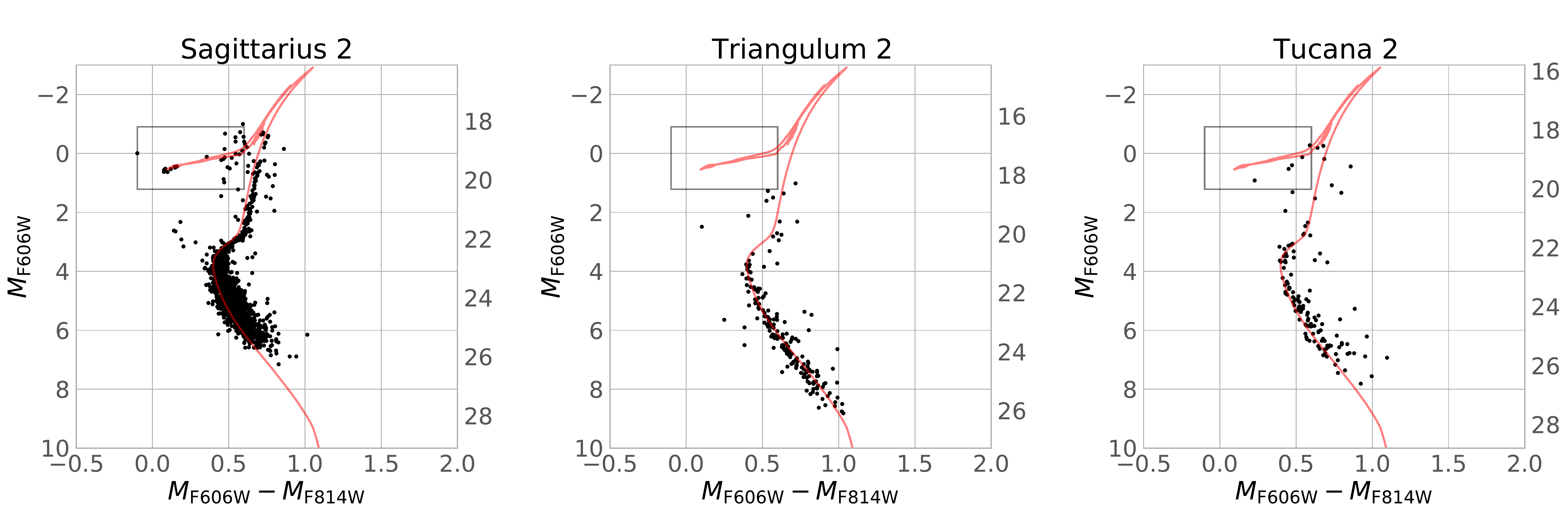}
\caption{Color-magnitude diagrams ($M_\mathrm{F606W}$ vs. $M_\mathrm{F606W}-M_\mathrm{F814W}$ in Vegamag) of the seven galaxies we analyzed. The gray box shows the approximate location of the HB feature. The red line is a reference isochrone with age $= 13.7$ Gyr and [Fe/H] = $-2.0$ from the MESA/MIST library \citep{Dotter2016}. The right y-axis shows the range of apparent magnitudes spanned by each galaxy. The foreground/background contamination will be accounted for in the CMD modeling phase.}
\label{cmds}
\end{figure*}

\section{Data and Photometry} \label{sec:phot}
Observations of a total of 30 UFDs were performed using the F606W and F814W filters of the \textit{HST} ACS Wide Field Channel (Treasury Program 14734; PI: Kallivayalil). The basic observing strategy included collecting four dithered 1100~s exposures in both filters for each target.

The images were processed through the current ACS pipeline, which corrects for charge-transfer inefficiency (CTI), and the separate dithers were combined using the DRIZZLE package to create the drc files. The CTI-corrected separate dither images, or flc files, were also used to create the photometric catalogs. The photutils routines \texttt{DAOStarFinder} and \texttt{aperture\_photometry} were used to detect sources and to calculate the flux inside circular apertures of both four- and six-pixel radii. 
We imposed two separate criteria to create a flag differentiating real sources from artifacts. First, we tracked whether a given source's magnitude in the four- and six-pixel radii was within 1.5 standard deviations of the median difference across all sources. The second criterion was that the magnitude difference between the two radii must be positive. Sources meeting both conditions were flagged as real.We performed encircled energy corrections and converted the flux to STMAG. The four-pixel radius drc magnitudes for each matching source between filters were used in the final photometric catalogs. Lastly, magnitudes were converted to the Vegamag photometric system.

Sources in the flc images underwent the same steps and were matched across the four separate dithers in each filter using a 6-parameter linear transformation. To derive an empirical error in the mean magnitude for the sources, we used the sigma-clipped standard deviation of the four-pixel radius flc magnitudes for each source in the separate filters. The flc sources were then matched between the filters before being matched to the drc sources using the same 6-parameter transformation method.

We analyze here the UFDs in the program with enough stars (at least 100) to perform a reliable CMD fit; for less populated CMDs, random uncertainties due to stochastic sampling of the CMD become critical and affect the reliability of the fitting technique.

We include Sag 2 in our sample, though there is conflicting evidence of it being either a UFD \citep{Longeard2020} or a globular cluster \citep{MutluPakdil2018,Longeard2021}; spectroscopic follow-up observations would be the ultimate confirmation of its true nature.

\section{Color-Magnitude Diagrams} \label{sec:cmds}
Figure \ref{cmds} presents the $M_\mathrm{F606W}$ vs. $M_\mathrm{F606W}-M_\mathrm{F814W}$ CMDs of the UFD galaxies analyzed here. We show absolute magnitudes, derived according to the appropriate distance and extinction (as in Table \ref{catalog}), to ease the comparison among galaxies (apparent magnitudes are on the right y-axis). The gray box shows the approximate location of the horizontal branch (HB), the core He-burning evolutionary phase of stars with masses $\lesssim 2$~M$_{\odot}$ and low metallicity ([Fe/H] $< -1.5$), while the red line is a reference isochrone with age $= 13.7$ Gyr and [Fe/H] = $-2.0$ from the MESA/MIST library \citep{Dotter2016}.

These CMDs are dominated by an ancient metal-poor population, and we can identify the oldest MS turnoff at $M_\mathrm{F606W} \sim 3$ and $M_\mathrm{F606W}-M_\mathrm{F814W} \sim 0.5$, which is our most sensitive and reliable age constraint. Using the isochrone as a guide, we also notice how the color, color spread, and turnoff morphology vary slightly from galaxy to galaxy, indicating their different SF and chemical evolution histories.

A few galaxies, in particular Hya~2, and Sag~2, show an extended HB, reaching very blue colors, while Hor~1 and Phx~2 may have a single blue HB candidate. The current empirical and theoretical evidence indicates that the HB morphology is affected by the initial metal content, with more metal-poor populations showing bluer HBs. However, there might be other, more complicated parameters affecting the HB morphology, and the modeling of this phase is still subject to great uncertainties (see, e.g. \citealt{Torelli2019}).

There are a few blue straggler stars in some of the CMDs, extending at bluer colors and brighter magnitudes than the dominant MS turnoff. This is particularly evident in Phx~2 and Sag~2. These stars are very common in ancient populations \citep{Santana2013}, but they can mimic a much younger sub-population. For example, the turnoff mass at $12-13$~Gyr is $\sim 0.8$~M$_{\odot}$, but blue stragglers can be even twice as massive, resembling a MS population of $\sim 2$~Gyr. This is why we exclude them from the CMD fitting (see Section \ref{sec:sfh}). \\

\section{Star Formation History Derivation} \label{sec:sfh}
To date, the most powerful approach to recover a SFH from an observed CMD is the comparison with models, a method applied by many different groups (see, e.g., the review by \citealt{Tolstoy2009}). Synthetic CMDs are built from a set of stellar models (evolutionary tracks or isochrones) adequately treated to match the distance, extinction, and photometric properties of the galaxy under analysis. To add the right scatter to the models, we adopt a scattering function based on fitting the photometric errors generated by the photometry pipeline (see Section \ref{sec:phot}).
Each synthetic CMD represents a simple stellar population of fixed age and metallicity, and a linear combination of these creates a composite population that can represent, with the appropriate weights, any SFH. The best-fitting weights are determined by using a minimization algorithm to compare data and models. To take into account the effect of field contamination, we include an additional component built from the Besan\c{c}on Galaxy model \citep{Robin2003} along the line of sight to each galaxy (as in, e.g., \citealt{Brown2014}).

We adopt Poisson maximum likelihood statistics to perform the minimization of the residuals between data and models, to accommodate the fact that some parts of the CMD might have a low number of stars. We implemented the construction of the synthetic CMDs and the comparison between models and data in the hybrid genetic code SFERA \citep{Cignoni2015}.

To take into account possible systematic effects due to the adopted stellar evolution models, we derive SFHs using two different sets of models, the \textit{Victoria-Regina} library \citep{VandenBerg2014} and the MESA/MIST library \citep{Dotter2016}. While the MESA/MIST models are available only for scaled-solar abundances, we adopt the \textit{Victoria-Regina} models with an enhancement of $+0.4$ for $\alpha$-elements, which is more appropriate for old metal-poor populations. While this should represent a good approximation for UFDs, some variations are observed from galaxy to galaxy, which might introduce some uncertainty in the ages estimates; however, we expect them to affect the SFHs at a much smaller level than the statistical and systematic uncertainties already taken into account in our analysis. Moreover, while various systematic uncertainties certainly affect our absolute age estimates, one of the main goals of this paper is to do a relative comparison between ages, which is more robust as systematics will affect all SFHs in the same way.

We use an isochrone grid with [Fe/H] in the range $[-1,-4]$ with 0.1 dex steps (using literature spectroscopic measurements as a further constraint), and ages in the range $[8, 13.7]$ Gyr with 100 Myr steps. We assume a \citet{Kroupa2001} IMF and $30\%$ of binaries drawn from the same IMF (\citealt{Spencer2017}; we checked that very little differences are found when different fractions or mass ratios are adopted). Distance moduli and extinctions are taken from the literature (Table \ref{catalog}), but the code allows for small variations in these parameters to maximize the likelihood.

\begin{figure*}
\centering
\raisebox{-0.4\height}{\includegraphics[height=0.3\linewidth]{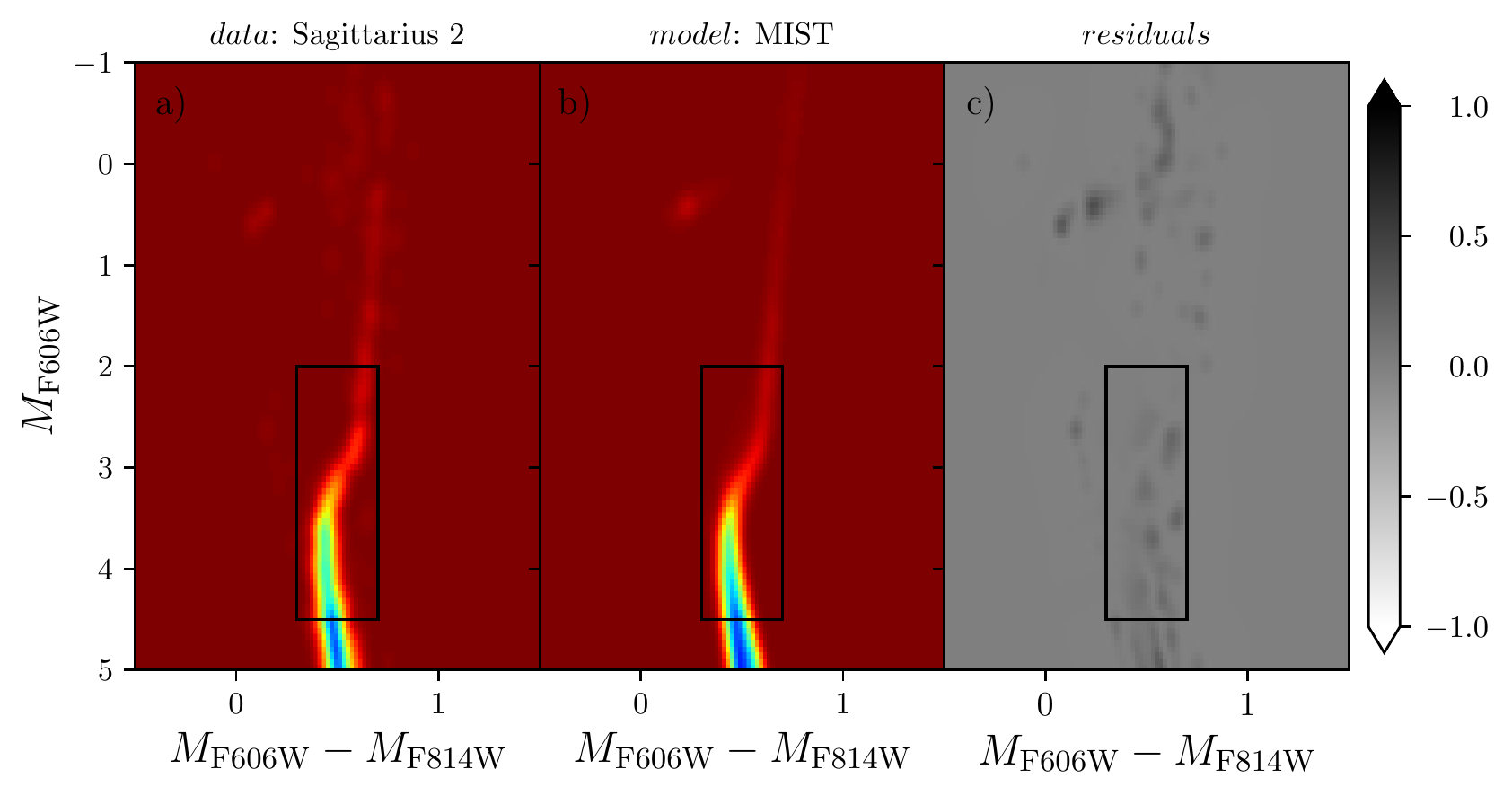}}
\raisebox{-0.36\height}{\includegraphics[height=0.3\linewidth]{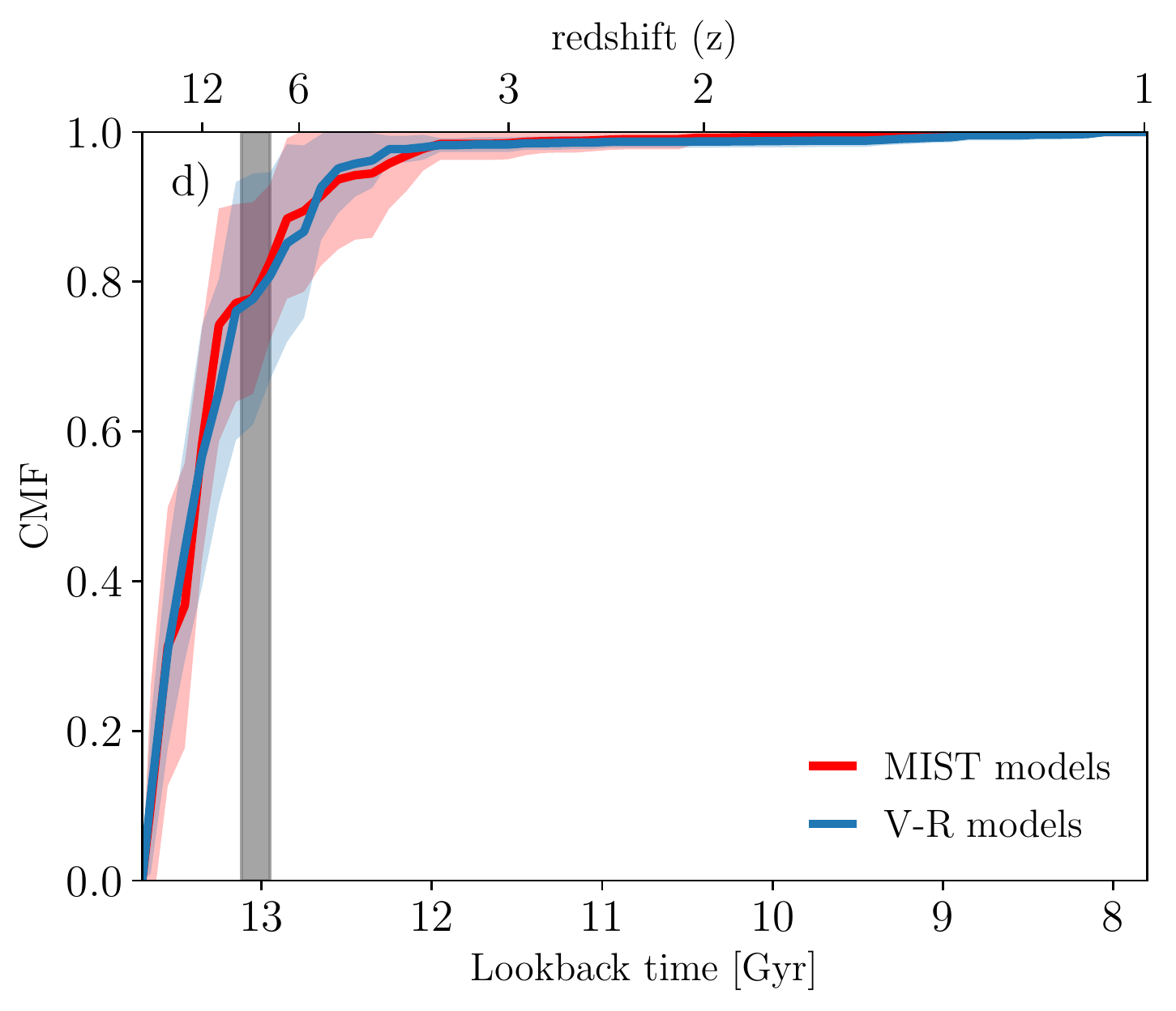}}
\caption{Example of the output from SFERA for the galaxy Sag~2. \textit{Panel a)}. Hess diagram of the $M_\mathrm{F606W}$ vs. $M_\mathrm{F606W}-M_\mathrm{F814W}$ CMD from our data. \textit{Panel b)}. Hess diagram reconstructed from the best-fit SFH on the basis of the MIST models. \textit{Panel c)}. Residuals between the two; the black box corresponds to the MS turnoff area, used for the fit. \textit{Panel d)}. Recovered cumulative SFHs (from the MIST models in red, from the \textit{Victoria-Regina} models in blue) with $1\sigma$ uncertainties; the gray band indicates the mid-point reionization
redshift $z_{\mathrm{re}}=7.7\pm0.7$ \citep{Plank2020}.}
\label{sfh}
\end{figure*}

We fit the CMD sequence from the MS turnoff to the top of the sub-giant branch (see the box in Figure \ref{sfh}), to focus on the region most sensitive to age variations, while at the same time avoiding areas which are poorly constrained by the models. Also, excluding the lower MS minimizes the sensitivity of the fit to the assumed IMF. Moreover, we avoid blue stragglers, which might mimic a much younger population as discussed in Section \ref{sec:cmds}. At the faintest end of the SFH derivation the observations are near 100\% complete, and the typical magnitude error is $0.01 - 0.02$ mag, depending on the galaxy.\\

\section{Results and Discussion}
An example of the output from SFERA is shown in Figure \ref{sfh} for Sag~2. Panels a) and b) show the Hess diagrams of the observed and recovered CMDs (the latter based on MIST models), while panel c) shows the residuals between the two; the black box corresponds to the MS turnoff area, used as a constraint for the SFH derivation. Panel d) shows the recovered cumulative SFHs (i.e., the fraction of total stellar mass formed prior to a given epoch) from both the MIST models (in red) and the \textit{Victoria-Regina} models (in blue); the gray band indicates the mid-point reionization redshift with its uncertainty, $z_{\mathrm{re}}=7.7\pm0.7$ \citep{Plank2020}.

Comparing observed and synthetic CMDs demonstrates how well we reproduce the data, even when fitting only the MS turnoff area. We can also reproduce the luminosity of the HB, even though the models do not reach the most extreme blue colors; this is one of the well-known complications in HB morphology modeling.

\begin{figure*}
\centering
\includegraphics[width=\linewidth]{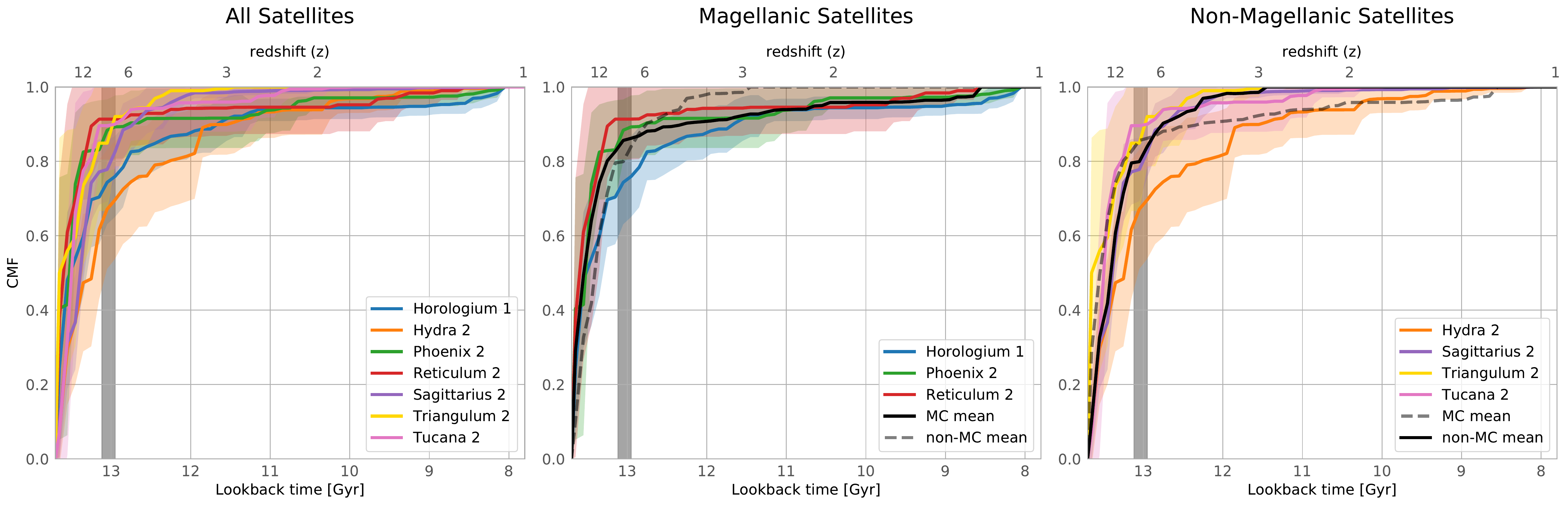}
\caption{\textit{Left panel}. Cumulative SFHs of the seven UFD galaxies we analyzed, with $1\sigma$ uncertainties. The gray band indicates the mid-point reionization
redshift $z_{\mathrm{re}}=7.7\pm0.7$ \citep{Plank2020}. \textit{Middle panel}. Same for the candidate Magellanic satellites; the black solid line shows the error-weighted average of the sub-sample, while the dashed gray line shows the error-weighted average of the other sub-sample (non-Magellanic satellites). \textit{Right panel}. Same for the non-Magellanic satellites.}
\label{all}
\end{figure*}

Figure \ref{all} (left panel) shows the cumulative SFHs for the galaxies we analyzed, with the same reference to reionization as in Figure \ref{sfh}. As expected, the SFHs are predominantly old, consistent with the galaxies having created at least 80\% of their stellar mass by $z\sim6$ (in agreement with, e.g., \citealt{Bose2018}). Table \ref{quenching} summarizes the time at which 50\% of the total stellar mass was formed ($\tau_{50}$) and the time at which 90\% of the total stellar mass was formed ($\tau_{90}$, quenching time, as in, e.g., \citealt{Weisz2015,Weisz2019}) by each galaxy. We adopt $\tau_{90}$ as a proxy for the quenching time to minimize possible residual contamination from blue stragglers and to ease the comparison of our results with the literature. All galaxies in our sample have quenching times older than 11.5 Gyr ago.

The middle and right panels of Figure \ref{all} show the same results divided in two sub-samples: candidate Magellanic satellites and non-Magellanic satellites, according to the results by \citet{Kallivayalil2018}, \citet{Erkal2019}, \citet{Fritz2019}, and \citet{Patel2020}. For each panel, the black solid curves also show the error-weighted average SFH of the two sub-samples, while the dashed gray line is the average of the other sub-sample. On average, we find $\tau_{90} = 12.06\pm0.72$ Gyr ago for Magellanic satellites and $\tau_{90} = 12.68\pm0.23$ Gyr ago for non-Magellanic satellites (this becomes $12.67 \pm 0.41$ if we exclude Sag 2); even though they are comparable within the errors, there is a marginal difference between these values (a two-sample KS test of the average distributions gives a p-value of $5.7\times10^{-10}$). Indeed, the candidate Magellanic satellites appear to have more prolonged SFHs compared to non-Magellanic ones, possibly due to the fact that they should have resided in low-density environments for most of their lifetimes; however, the uncertainties are too large to reach a definitive conclusion.

Another interesting consideration is that, among the non-Magellanic satellites, the one with the youngest SFH is Hya~2, i.e., the most distant one in the sample ($\sim 150$~kpc, whereas all others lie within $\sim 80$~kpc), and the most massive one, which is also likely on first infall into the MW’s halo \citep{Patel2020}. However, distance makes the SFH uncertainties quite large, so these results should be regarded with caution. The shape of the SFH is also different for the two sub-samples, with non-Magellanic satellites reaching 100\% of the stellar mass assembly within 11.5 Gyr ago, i.e. $z \sim 3$, while MC satellites show more prolonged star formation and reach the same point only 8.5 Gyr ago.

Similarly, in a study comparing the quenching times of M31 and MW satellites, \citet{Weisz2019} found that the two populations do not share many trends, though the authors do not have measurements for similar-mass UFDs in M31. Despite the uncertainties (due to the limited sample of UFDs around M31 and the fact that the satellites' distances prevent reaching the oldest MS turnoff), they suggest that a connection between the SFHs of satellites and their host galaxy's accretion history could be the best explanation for this different behavior.

These results, together with the extended SFHs we find for MC satellites, might support the idea that satellites of low-mass hosts experience a weaker ionization field, resulting in a more prolonged SFH than if they were satellites of more massive hosts, as found by, e.g., \citet{Joshi2021}. On the other hand, it is important to be mindful of our uncertainties, and be cautious about over-interpreting these results. Within the error bars, all galaxies in our sample are compatible with being quenched by reionization, as found in other works \citep[e.g.,][]{Brown2014,Weisz2015,Tollerud2018,Fillingham2019}.

From the numerical point of view, simulations reaching the necessary resolution for UFDs have mainly examined field galaxies \citep[e.g.,][]{Jeon2017,Wheeler2019}, and find that reionization and SNe feedback are key factors to their quenching. More recently, \citet{Applebaum2021} presented a suite of cosmological simulations that contain zoom-ins of MW-like galaxies, with simulated satellites extending to the UFD regime. By comparing UFD satellites to UFDs in the field, the authors found that reionization and feedback were indeed the main quenching mechanisms, rather than environmental effects (it is worth noting that these simulations lack radiative feedback from the host galaxies and adopt a uniform reionization model). However, studies examining the SFHs of UFDs around LMC-like dwarfs are not available, yet. \\

%%% nice but not necessary
% Even with all the proper caution and caveats, it is very interesting to see that differences in SF shape and quenching behavior are detectable in our SFHs, as this could be a starting point to use SFHs as an additional check regarding satellite associations and environmental effects on these small systems, in particular with a larger sample and more precise constraints from imaging and spectroscopic follow-ups. \\

\begin{table}[t]
\caption{Summary of the time at which 50\% of the total stellar was mass formed ($\tau_{50}$) and the time at which 90\% of the total stellar mass was formed ($\tau_{90}$, quenching time) for each galaxy. The list is sorted by quenching time (from oldest to youngest). The last column indicates whether a galaxy is a potential Magellanic satellite \citep{Erkal2019,Patel2020}.}
\begin{center}
\begin{tabular}{lccc}
    \toprule
    \midrule
    \addlinespace[0.3em]
    Galaxy & $\tau_{50}$ [Gyr ago] & $\tau_{90}$ [Gyr ago] & MC satellite? \\
    \midrule
    Tri 2 & $13.50\pm0.23$ & $12.91\pm0.53$ & \\
    Tuc 2 & $13.42\pm0.17$ & $12.84\pm0.84$ & \\
    Sag 2 & $13.40\pm0.07$ & $12.68\pm0.28$ & \\
    Phx 2 & $13.50\pm0.14$ & $12.47\pm1.10$ & yes \\
    Ret 2 & $13.52\pm0.16$ & $12.29\pm1.77$ & yes \\
    Hya 2 & $13.21\pm0.32$ & $11.59\pm1.00$ & \\
    Hor 1 & $13.44\pm0.17$ & $11.53\pm1.13$ & yes \\
    \midrule
    \bottomrule
\end{tabular}
\end{center}
\label{quenching}
\end{table}

Despite the challenges, studies of the kind presented here are fundamental to fully explore the properties of low-mass galaxies, and to understand the impact of environment and reionization on such systems. A detailed analysis and comparison of the SFHs of UFDs residing in different environments at early times is also a way to discover variations in the ionization field over large scales. The observations presented here will also be an important baseline for follow-up imaging with \textit{HST} and the \textit{James Webb Space Telescope}, to measure both bulk PMs and internal motions, while spectroscopic surveys will improve metallicity constraints (thus allowing also more precise age derivations) on these very interesting galaxies.

%TC:ignore
\acknowledgments{
These data are associated with the \textit{HST} Treasury Program 14734 (PI Kallivayalil). Support for this program was provided by NASA through grants from the Space Telescope Science Institute. AW received support from NASA ATP grants 80NSSC18K1097 and 80NSSC20K0513; a Scialog Award from the Heising-Simons Foundation; and a Hellman Fellowship. This material is based upon work supported by the National Science Foundation under grant No. AST-1847909.  ENK gratefully acknowledges support from a Cottrell Scholar award administered by the Research Corporation for Science Advancement.
}
%TC:endignore

\bibliography{bib.bib}

\begin{thebibliography}{}
\expandafter\ifx\csname natexlab\endcsname\relax\def\natexlab#1{#1}\fi

\bibitem[{{Applebaum} {et~al.}(2021){Applebaum}, {Brooks}, {Christensen},
  {Munshi}, {Quinn}, {Shen}, \& {Tremmel}}]{Applebaum2021}
{Applebaum}, E., {Brooks}, A.~M., {Christensen}, C.~R., {et~al.} 2021, \apj,
  906, 96

\bibitem[{{Bechtol} {et~al.}(2015){Bechtol}, {Drlica-Wagner}, {Balbinot},
  {Pieres}, {Simon}, {Yanny}, \& {et al.}}]{Bechtol2015}
{Bechtol}, K., {Drlica-Wagner}, A., {Balbinot}, E., {et~al.} 2015, \apj, 807,
  50

\bibitem[{{Besla} {et~al.}(2018){Besla}, {Patton}, {Stierwalt},
  {Rodriguez-Gomez}, {Patel}, {Kallivayalil}, {Johnson}, {Pearson}, {Privon},
  \& {Putman}}]{Besla2018}
{Besla}, G., {Patton}, D.~R., {Stierwalt}, S., {et~al.} 2018, \mnras, 480, 3376

\bibitem[{{Bose} {et~al.}(2018){Bose}, {Deason}, \& {Frenk}}]{Bose2018}
{Bose}, S., {Deason}, A.~J., \& {Frenk}, C.~S. 2018, \apj, 863, 123

\bibitem[{{Bovill} \& {Ricotti}(2009)}]{Bovill2009}
{Bovill}, M.~S., \& {Ricotti}, M. 2009, \apj, 693, 1859

\bibitem[{{Brown} {et~al.}(2014){Brown}, {Tumlinson}, {Geha}, {Simon},
  {Vargas}, {VandenBerg}, {Kirby}, {Kalirai}, {Avila}, {Gennaro}, {Ferguson},
  {Mu{\~n}oz}, {Guhathakurta}, \& {Renzini}}]{Brown2014}
{Brown}, T.~M., {Tumlinson}, J., {Geha}, M., {et~al.} 2014, \apj, 796, 91

\bibitem[{{Cignoni} {et~al.}(2015){Cignoni}, {Sabbi}, {van der Marel}, {Tosi},
  {Zaritsky}, {Anderson}, {Lennon}, {Aloisi}, {de Marchi}, {Gouliermis},
  {Grebel}, {Smith}, \& {Zeidler}}]{Cignoni2015}
{Cignoni}, M., {Sabbi}, E., {van der Marel}, R.~P., {et~al.} 2015, \apj, 811,
  76

\bibitem[{{Deason} {et~al.}(2015){Deason}, {Wetzel}, {Garrison-Kimmel}, \&
  {Belokurov}}]{Deason2015}
{Deason}, A.~J., {Wetzel}, A.~R., {Garrison-Kimmel}, S., \& {Belokurov}, V.
  2015, \mnras, 453, 3568

\bibitem[{{D'Onghia} \& {Lake}(2008)}]{Donghia2008}
{D'Onghia}, E., \& {Lake}, G. 2008, \apjl, 686, L61

\bibitem[{{Dooley} {et~al.}(2017){Dooley}, {Peter}, {Carlin}, {Frebel},
  {Bechtol}, \& {Willman}}]{Dooley2017}
{Dooley}, G.~A., {Peter}, A. H.~G., {Carlin}, J.~L., {et~al.} 2017, \mnras,
  472, 1060

\bibitem[{{Dotter}(2016)}]{Dotter2016}
{Dotter}, A. 2016, \apjs, 222, 8

\bibitem[{{Drlica-Wagner} {et~al.}(2015){Drlica-Wagner}, {Bechtol}, {Rykoff},
  {Luque}, {Queiroz}, {Mao}, {Wechsler}, {Simon}, {Santiago}, {Yanny},
  {Balbinot}, {Dodelson}, {Fausti Neto}, {James}, {Li}, {Maia}, {Marshall},
  {Pieres}, {Stringer}, {Walker}, {Abbott}, {Abdalla}, {Allam},
  {Benoit-L{\'e}vy}, {Bernstein}, {Bertin}, {Brooks}, {Buckley-Geer}, {Burke},
  {Carnero Rosell}, {Carrasco Kind}, {Carretero}, {Crocce}, {da Costa},
  {Desai}, {Diehl}, {Dietrich}, {Doel}, {Eifler}, {Evrard}, {Finley},
  {Flaugher}, {Fosalba}, {Frieman}, {Gaztanaga}, {Gerdes}, {Gruen}, {Gruendl},
  {Gutierrez}, {Honscheid}, {Kuehn}, {Kuropatkin}, {Lahav}, {Martini},
  {Miquel}, {Nord}, {Ogando}, {Plazas}, {Reil}, {Roodman}, {Sako}, {Sanchez},
  {Scarpine}, {Schubnell}, {Sevilla-Noarbe}, {Smith}, {Soares-Santos},
  {Sobreira}, {Suchyta}, {Swanson}, {Tarle}, {Tucker}, {Vikram}, {Wester},
  {Zhang}, {Zuntz}, \& {DES Collaboration}}]{Drlica-Wagner2015}
{Drlica-Wagner}, A., {Bechtol}, K., {Rykoff}, E.~S., {et~al.} 2015, \apj, 813,
  109

\bibitem[{{Erkal} \& {Belokurov}(2019)}]{Erkal2019}
{Erkal}, D., \& {Belokurov}, V.~A. 2019, arXiv e-prints, arXiv:1907.09484

\bibitem[{{Fillingham} {et~al.}(2019){Fillingham}, {Cooper}, {Kelley},
  {Rodriguez Wimberly}, {Boylan-Kolchin}, {Bullock}, {Garrison-Kimmel},
  {Pawlowski}, \& {Wheeler}}]{Fillingham2019}
{Fillingham}, S.~P., {Cooper}, M.~C., {Kelley}, T., {et~al.} 2019, arXiv
  e-prints, arXiv:1906.04180

\bibitem[{{Fritz} {et~al.}(2018){Fritz}, {Battaglia}, {Pawlowski},
  {Kallivayalil}, {van der Marel}, {Sohn}, {Brook}, \& {Besla}}]{Fritz2018}
{Fritz}, T.~K., {Battaglia}, G., {Pawlowski}, M.~S., {et~al.} 2018, \aap, 619,
  A103

\bibitem[{{Fritz} {et~al.}(2019){Fritz}, {Carrera}, {Battaglia}, \&
  {Taibi}}]{Fritz2019}
{Fritz}, T.~K., {Carrera}, R., {Battaglia}, G., \& {Taibi}, S. 2019, \aap, 623,
  A129

\bibitem[{{Jahn} {et~al.}(2019){Jahn}, {Sales}, {Wetzel}, {Boylan-Kolchin},
  {Chan}, {El-Badry}, {Lazar}, \& {Bullock}}]{Jahn2019}
{Jahn}, E.~D., {Sales}, L.~V., {Wetzel}, A., {et~al.} 2019, \mnras, 489, 5348

\bibitem[{{Jeon} {et~al.}(2017){Jeon}, {Besla}, \& {Bromm}}]{Jeon2017}
{Jeon}, M., {Besla}, G., \& {Bromm}, V. 2017, \apj, 848, 85

\bibitem[{{Jeon} {et~al.}(2021){Jeon}, {Bromm}, {Besla}, {Yoon}, \&
  {Choi}}]{Jeon2021}
{Jeon}, M., {Bromm}, V., {Besla}, G., {Yoon}, J., \& {Choi}, Y. 2021, \mnras,
  502, 1

\bibitem[{{Joshi} {et~al.}(2021){Joshi}, {Pillepich}, {Nelson}, {Zinger},
  {Marinacci}, {Springel}, {Vogelsberger}, \& {Hernquist}}]{Joshi2021}
{Joshi}, G.~D., {Pillepich}, A., {Nelson}, D., {et~al.} 2021, arXiv e-prints,
  arXiv:2101.12226

\bibitem[{{Kallivayalil} {et~al.}(2018){Kallivayalil}, {Sales}, {Zivick},
  {Fritz}, {Del Pino}, {Sohn}, {Besla}, {van der Marel}, {Navarro}, \&
  {Sacchi}}]{Kallivayalil2018}
{Kallivayalil}, N., {Sales}, L.~V., {Zivick}, P., {et~al.} 2018, \apj, 867, 19

\bibitem[{{Koposov} {et~al.}(2015){Koposov}, {Belokurov}, {Torrealba}, \&
  {Evans}}]{Koposov2015}
{Koposov}, S.~E., {Belokurov}, V., {Torrealba}, G., \& {Evans}, N.~W. 2015,
  \apj, 805, 130

\bibitem[{{Koposov} {et~al.}(2018){Koposov}, {Walker}, {Belokurov}, {Casey},
  {Geringer-Sameth}, {Mackey}, {Da Costa}, {Erkal}, {Jethwa}, {Mateo},
  {Olszewski}, \& {Bailey}}]{Koposov2018}
{Koposov}, S.~E., {Walker}, M.~G., {Belokurov}, V., {et~al.} 2018, \mnras, 479,
  5343

\bibitem[{{Kroupa}(2001)}]{Kroupa2001}
{Kroupa}, P. 2001, \mnras, 322, 231

\bibitem[{{Laevens} {et~al.}(2015){Laevens}, {Martin}, {Bernard}, \& {et
  al.}}]{Laevens2015}
{Laevens}, B.~P.~M., {Martin}, N.~F., {Bernard}, E.~J., \& {et al.} 2015, \apj,
  813, 44

\bibitem[{{Longeard} {et~al.}(2020){Longeard}, {Martin}, {Starkenburg},
  {Ibata}, {Collins}, {Laevens}, {Mackey}, {Rich}, {Aguado}, {Arentsen},
  {Jablonka}, {Gonz{\'a}lez Hern{\'a}ndez}, {Navarro}, \&
  {S{\'a}nchez-Janssen}}]{Longeard2020}
{Longeard}, N., {Martin}, N., {Starkenburg}, E., {et~al.} 2020, \mnras, 491,
  356

\bibitem[{{Longeard} {et~al.}(2021){Longeard}, {Martin}, {Ibata},
  {Starkenburg}, {Jablonka}, {Aguado}, {Carlberg}, {C{\^o}t{\'e}},
  {Gonz{\'a}lez Hern{\'a}ndez}, {Lucchesi}, {Malhan}, {Navarro},
  {S{\'a}nchez-Janssen}, {Thomas}, {Venn}, \& {McConnachie}}]{Longeard2021}
{Longeard}, N., {Martin}, N., {Ibata}, R.~A., {et~al.} 2021, \mnras, 503, 2754

\bibitem[{{Martin} {et~al.}(2015){Martin}, {Nidever}, {Besla}, {Olsen},
  {Walker}, {Vivas}, {Gruendl}, {Kaleida}, {Mu{\~n}oz}, {Blum}, {Saha}, {Conn},
  {Bell}, {Chu}, {Cioni}, {de Boer}, {Gallart}, {Jin}, {Kunder}, {Majewski},
  {Martinez-Delgado}, {Monachesi}, {Monelli}, {Monteagudo}, {No{\"e}l},
  {Olszewski}, {Stringfellow}, {van der Marel}, \& {Zaritsky}}]{Martin2015}
{Martin}, N.~F., {Nidever}, D.~L., {Besla}, G., {et~al.} 2015, \apjl, 804, L5

\bibitem[{{Mutlu-Pakdil} {et~al.}(2018){Mutlu-Pakdil}, {Sand}, {Carlin},
  {Spekkens}, {Caldwell}, {Crnojevi{\'c}}, {Hughes}, {Willman}, \&
  {Zaritsky}}]{MutluPakdil2018}
{Mutlu-Pakdil}, B., {Sand}, D.~J., {Carlin}, J.~L., {et~al.} 2018, \apj, 863,
  25

\bibitem[{{Nidever} {et~al.}(2017){Nidever}, {Olsen}, {Walker}, {Vivas},
  {Blum}, {Kaleida}, {Choi}, {Conn}, {Gruendl}, {Bell}, {Besla}, {Mu{\~n}oz},
  {Gallart}, {Martin}, {Olszewski}, {Saha}, {Monachesi}, {Monelli}, {de Boer},
  {Johnson}, {Zaritsky}, {Stringfellow}, {van der Marel}, {Cioni}, {Jin},
  {Majewski}, {Martinez-Delgado}, {Monteagudo}, {No{\"e}l}, {Bernard},
  {Kunder}, {Chu}, {Bell}, {Santana}, {Frechem}, {Medina}, {Parkash},
  {Navarrete}, \& {Hayes}}]{Nidever2017}
{Nidever}, D.~L., {Olsen}, K., {Walker}, A.~R., {et~al.} 2017, \aj, 154, 199

\bibitem[{{Patel} {et~al.}(2020){Patel}, {Kallivayalil}, {Garavito-Camargo},
  {Besla}, {Weisz}, {van der Marel}, {Boylan-Kolchin}, {Pawlowski}, \&
  {G{\'o}mez}}]{Patel2020}
{Patel}, E., {Kallivayalil}, N., {Garavito-Camargo}, N., {et~al.} 2020, arXiv
  e-prints, arXiv:2001.01746

\bibitem[{{Planck Collaboration} {et~al.}(2020){Planck Collaboration},
  {Aghanim}, {Akrami}, {Ashdown}, {Aumont}, {Baccigalupi}, {Ballardini},
  {Banday}, {Barreiro}, {Bartolo}, {Basak}, {Battye}, {Benabed}, {Bernard},
  {Bersanelli}, {Bielewicz}, {Bock}, {Bond}, {Borrill}, {Bouchet}, {Boulanger},
  {Bucher}, {Burigana}, {Butler}, {Calabrese}, {Cardoso}, {Carron},
  {Challinor}, {Chiang}, {Chluba}, {Colombo}, {Combet}, {Contreras}, {Crill},
  {Cuttaia}, {de Bernardis}, {de Zotti}, {Delabrouille}, {Delouis}, {Di
  Valentino}, {Diego}, {Dor{\'e}}, {Douspis}, {Ducout}, {Dupac}, {Dusini},
  {Efstathiou}, {Elsner}, {En{\ss}lin}, {Eriksen}, {Fantaye}, {Farhang},
  {Fergusson}, {Fernandez-Cobos}, {Finelli}, {Forastieri}, {Frailis},
  {Fraisse}, {Franceschi}, {Frolov}, {Galeotta}, {Galli}, {Ganga},
  {G{\'e}nova-Santos}, {Gerbino}, {Ghosh}, {Gonz{\'a}lez-Nuevo}, {G{\'o}rski},
  {Gratton}, {Gruppuso}, {Gudmundsson}, {Hamann}, {Handley}, {Hansen},
  {Herranz}, {Hildebrandt}, {Hivon}, {Huang}, {Jaffe}, {Jones}, {Karakci},
  {Keih{\"a}nen}, {Keskitalo}, {Kiiveri}, {Kim}, {Kisner}, {Knox},
  {Krachmalnicoff}, {Kunz}, {Kurki-Suonio}, {Lagache}, {Lamarre}, {Lasenby},
  {Lattanzi}, {Lawrence}, {Le Jeune}, {Lemos}, {Lesgourgues}, {Levrier},
  {Lewis}, {Liguori}, {Lilje}, {Lilley}, {Lindholm}, {L{\'o}pez-Caniego},
  {Lubin}, {Ma}, {Mac{\'\i}as-P{\'e}rez}, {Maggio}, {Maino}, {Mandolesi},
  {Mangilli}, {Marcos-Caballero}, {Maris}, {Martin}, {Martinelli},
  {Mart{\'\i}nez-Gonz{\'a}lez}, {Matarrese}, {Mauri}, {McEwen}, {Meinhold},
  {Melchiorri}, {Mennella}, {Migliaccio}, {Millea}, {Mitra},
  {Miville-Desch{\^e}nes}, {Molinari}, {Montier}, {Morgante}, {Moss}, {Natoli},
  {N{\o}rgaard-Nielsen}, {Pagano}, {Paoletti}, {Partridge}, {Patanchon},
  {Peiris}, {Perrotta}, {Pettorino}, {Piacentini}, {Polastri}, {Polenta},
  {Puget}, {Rachen}, {Reinecke}, {Remazeilles}, {Renzi}, {Rocha}, {Rosset},
  {Roudier}, {Rubi{\~n}o-Mart{\'\i}n}, {Ruiz-Granados}, {Salvati}, {Sandri},
  {Savelainen}, {Scott}, {Shellard}, {Sirignano}, {Sirri}, {Spencer},
  {Sunyaev}, {Suur-Uski}, {Tauber}, {Tavagnacco}, {Tenti}, {Toffolatti},
  {Tomasi}, {Trombetti}, {Valenziano}, {Valiviita}, {Van Tent}, {Vibert},
  {Vielva}, {Villa}, {Vittorio}, {Wandelt}, {Wehus}, {White}, {White},
  {Zacchei}, \& {Zonca}}]{Plank2020}
{Planck Collaboration}, {Aghanim}, N., {Akrami}, Y., {et~al.} 2020, \aap, 641,
  A6

\bibitem[{{Robin} {et~al.}(2003){Robin}, {Reyl{\'e}}, {Derri{\`e}re}, \&
  {Picaud}}]{Robin2003}
{Robin}, A.~C., {Reyl{\'e}}, C., {Derri{\`e}re}, S., \& {Picaud}, S. 2003,
  \aap, 409, 523

\bibitem[{{Sales} {et~al.}(2011){Sales}, {Navarro}, {Cooper}, {White}, {Frenk},
  \& {Helmi}}]{Sales2011}
{Sales}, L.~V., {Navarro}, J.~F., {Cooper}, A.~P., {et~al.} 2011, \mnras, 418,
  648

\bibitem[{{Sales} {et~al.}(2017){Sales}, {Navarro}, {Kallivayalil}, \&
  {Frenk}}]{Sales2017}
{Sales}, L.~V., {Navarro}, J.~F., {Kallivayalil}, N., \& {Frenk}, C.~S. 2017,
  \mnras, 465, 1879

\bibitem[{{Santana} {et~al.}(2013){Santana}, {Mu{\~n}oz}, {Geha},
  {C{\^o}t{\'e}}, {Stetson}, {Simon}, \& {Djorgovski}}]{Santana2013}
{Santana}, F.~A., {Mu{\~n}oz}, R.~R., {Geha}, M., {et~al.} 2013, \apj, 774, 106

\bibitem[{{Simon}(2019)}]{Simon2019}
{Simon}, J.~D. 2019, \araa, 57, 375

\bibitem[{{Spencer} {et~al.}(2017){Spencer}, {Mateo}, {Walker}, {Olszewski},
  {McConnachie}, {Kirby}, \& {Koch}}]{Spencer2017}
{Spencer}, M.~E., {Mateo}, M., {Walker}, M.~G., {et~al.} 2017, \aj, 153, 254

\bibitem[{{Tollerud} \& {Peek}(2018)}]{Tollerud2018}
{Tollerud}, E.~J., \& {Peek}, J.~E.~G. 2018, \apj, 857, 45

\bibitem[{{Tolstoy} {et~al.}(2009){Tolstoy}, {Hill}, \& {Tosi}}]{Tolstoy2009}
{Tolstoy}, E., {Hill}, V., \& {Tosi}, M. 2009, \araa, 47, 371

\bibitem[{{Torelli} {et~al.}(2019){Torelli}, {Iannicola}, {Stetson}, {Ferraro},
  {Bono}, {Salaris}, {Castellani}, {Dall'Ora}, {Fontana}, {Monelli}, \&
  {Pietrinferni}}]{Torelli2019}
{Torelli}, M., {Iannicola}, G., {Stetson}, P.~B., {et~al.} 2019, \aap, 629, A53

\bibitem[{{Torrealba} {et~al.}(2016){Torrealba}, {Koposov}, {Belokurov}, \&
  {Irwin}}]{Torrealba2016}
{Torrealba}, G., {Koposov}, S.~E., {Belokurov}, V., \& {Irwin}, M. 2016,
  \mnras, 459, 2370

\bibitem[{{Torrealba} {et~al.}(2018){Torrealba}, {Belokurov}, {Koposov},
  {Bechtol}, {Drlica-Wagner}, {Olsen}, {Vivas}, {Yanny}, {Jethwa}, {Walker},
  {Li}, {Allam}, {Conn}, {Gallart}, {Gruendl}, {James}, {Johnson}, {Kuehn},
  {Kuropatkin}, {Martin}, {Martinez-Delgado}, {Nidever}, {No{\"e}l}, {Simon},
  {Stringfellow}, \& {Tucker}}]{Torrealba2018}
{Torrealba}, G., {Belokurov}, V., {Koposov}, S.~E., {et~al.} 2018, \mnras, 475,
  5085

\bibitem[{{VandenBerg} {et~al.}(2014){VandenBerg}, {Bergbusch}, {Ferguson}, \&
  {Edvardsson}}]{VandenBerg2014}
{VandenBerg}, D.~A., {Bergbusch}, P.~A., {Ferguson}, J.~W., \& {Edvardsson}, B.
  2014, \apj, 794, 72

\bibitem[{{Wang} {et~al.}(2020){Wang}, {Bose}, {Frenk}, {Gao}, {Jenkins},
  {Springel}, \& {White}}]{Wang2020}
{Wang}, J., {Bose}, S., {Frenk}, C.~S., {et~al.} 2020, \nat, 585, 39

\bibitem[{{Weisz} {et~al.}(2014){Weisz}, {Dolphin}, {Skillman}, {Holtzman},
  {Gilbert}, {Dalcanton}, \& {Williams}}]{Weisz2014}
{Weisz}, D.~R., {Dolphin}, A.~E., {Skillman}, E.~D., {et~al.} 2014, \apj, 789,
  148

\bibitem[{{Weisz} {et~al.}(2015){Weisz}, {Dolphin}, {Skillman}, {Holtzman},
  {Gilbert}, {Dalcanton}, \& {Williams}}]{Weisz2015}
{Weisz}, D.~R., {Dolphin}, A.~E., {Skillman}, E.~D., {et~al.} 2015, \apj, 804,
  136

\bibitem[{{Weisz} {et~al.}(2019){Weisz}, {Martin}, {Dolphin}, {Albers},
  {Collins}, {Ferguson}, {Lewis}, {Mackey}, {McConnachie}, {Rich}, \&
  {Skillman}}]{Weisz2019}
{Weisz}, D.~R., {Martin}, N.~F., {Dolphin}, A.~E., {et~al.} 2019, \apjl, 885,
  L8

\bibitem[{{Wetzel} {et~al.}(2016){Wetzel}, {Hopkins}, {Kim},
  {Faucher-Gigu{\`e}re}, {Kere{\v{s}}}, \& {Quataert}}]{Wetzel2016}
{Wetzel}, A.~R., {Hopkins}, P.~F., {Kim}, J.-h., {et~al.} 2016, \apjl, 827, L23

\bibitem[{{Wheeler} {et~al.}(2019){Wheeler}, {Hopkins}, {Pace},
  {Garrison-Kimmel}, {Boylan-Kolchin}, {Wetzel}, {Bullock}, {Kere{\v{s}}},
  {Faucher-Gigu{\`e}re}, \& {Quataert}}]{Wheeler2019}
{Wheeler}, C., {Hopkins}, P.~F., {Pace}, A.~B., {et~al.} 2019, \mnras, 490,
  4447

\end{thebibliography}

\end{document}